\pdfoutput=1 
\documentclass[11pt]{article}

\usepackage[T1]{fontenc}
\usepackage[cp1250]{inputenc}
\usepackage{graphicx}
\usepackage{psfrag}
\usepackage{color} 
\usepackage{bbm}
\usepackage{amsmath}
\usepackage{hyperref}
\usepackage{pdfpages}

\hypersetup{pdfstartview={FitH}}

\newcommand{\bq}{\begin{equation}}
\newcommand{\eq}{\end{equation}}
\newcommand{\ba}{\begin{eqnarray}}
\newcommand{\ea}{\end{eqnarray}}

\begin{document} 



# Spatial model predicts dispersal and cell turnover cause reduced intra-tumor heterogeneity

Bartlomiej Waclaw[1], Ivana Bozic[2,3], Meredith E. Pittman[4], Ralph H. Hruban[4], Bert Vogelstein[4,5], Martin A. Nowak[2,3,6]

[1]School of Physics and Astronomy, University of Edinburgh, Edinburgh, UK

[2]Program for Evolutionary Dynamics, Harvard University, Cambridge, USA

[3]Department of Mathematics, Harvard University, Cambridge, USA

[4]The Sol Goldman Pancreatic Cancer Research Center, Department of Pathology, Johns Hopkins University School of Medicine, Baltimore, USA

[5]Ludwig Center and Howard Hughes Medical Institute, Johns Hopkins Kimmel Cancer Center, Baltimore, USA

[6]Department of Organismic and Evolutionary Biology, Harvard University, Cambridge, USA

## Abstract

Most cancers in humans are large, measuring centimeters in diameter, composed of many billions of cells. An equivalent mass of normal cells would be highly heterogeneous as a result of the mutations that occur during each cell division. What is remarkable about cancers is their homogeneity - virtually every neoplastic cell within a large cancer contains the same core set of genetic alterations, with heterogeneity confined to mutations that have emerged after the last clonal expansions. How such clones expand within the spatially-constrained three dimensional architecture of a tumor, and come to dominate a large, pre-existing lesion, has never been explained. We here describe a model for tumor evolution that shows how short-range migration and cell turnover can account for rapid cell mixing inside the tumor. With it, we show that even a small selective advantage of a single cell within a large tumor allows the descendants of that cell to replace the precursor mass in a clinically relevant time frame. We also demonstrate that the same mechanisms can be responsible for the rapid onset of resistance to chemotherapy. Our model not only provides novel insights into spatial and temporal aspects of tumor growth but also suggests that targeting short range cellular migratory activity could have dramatic effects on tumor growth rates.

## 1. Introduction

Tumor growth is initiated when a single cell acquires genetic or epigenetic alterations that change the cell's net growth rate (birth minus death) and enable its progeny to outgrow surrounding cells. As these small lesions grow, the cells acquire additional alterations that cause them to multiply even faster and to change their metabolism to better survive the harsh conditions and nutrient deprivation. This progression eventually leads to a malignant tumor that can invade surrounding tissues and spread to other organs. Typical solid tumors contain about 30-70 clonal amino-acid-changing mutations that have accumulated during this multi-stage progression [1]. Despite the fact that a majority of these mutations are believed to be passengers that do not affect growth, and only ~5% to 10% are drivers that provide cells with a small selective growth advantage, a major fraction of the mutations,

particularly the drivers, are present in 30%-80% of cells in the primary tumor as well as in metastatic lesions derived from it [2,3,4,5].

Most attempts at explaining the genetic make-up of tumors assume well-mixed populations of cells and do not incorporate spatial constraints [6,7,8,9,10,11]. Several models of the genetic evolution of expanding tumors have been developed in the past [12,13,14,15,16,17,18], but they either assume very few mutations [12,13,14,15,16] or one- or two-dimensional growth [15,16,17,18]. On the other hand, models that incorporate spatial limitations have been developed to help understand processes such as tumor metabolism [19,20], angiogenesis [21,22,23], and cell migration [24], but these models ignore genetics. Here, we formulate a model that combines spatial growth and genetic evolution and use the model to describe the growth of primary tumors and metastases, as well as the development of resistance to therapeutic agents.

## 2. Metastasis

We first model the expansion of a metastatic lesion derived from a cancer cell that has escaped its primary site (e.g., breast or colorectal epithelium) and travelled through the circulation until it lodged at a distant site (e.g., lung or liver). The cell initiating the metastatic lesion is assumed to have all the driver gene mutations needed to expand. Motivated by histopathological images (Fig. 1a) we model the lesion as a conglomerate of "balls" of cells. Cells occupy sites in a regular three-dimensional lattice (Fig. 6a-b). Cells replicate stochastically with rates proportional to the number of surrounding empty sites (non-neoplastic cells or extracellular matrix), hence replication is faster at the edge of the tumor. This is supported by experimental data (Fig. 1b,c and Table 1). A cell with no cancer cell neighbors replicates at the maximal rate of $b=\ln(2)=0.69$ days$^{-1}$, equivalent to 24h cell doubling time, and a cell that is completely surrounded by other cancer cells does not replicate. Cells can also mutate, but we assume all mutations are passengers (they do not confer fitness advantages). Upon replication, a cell moves with a small probability $M$ to a nearby place close to the surface of the lesion and creates a new lesion. This "sprouting" of initial lesions can be due to short-range migration following an epithelial-to-mesenchymal transition[25] and consecutive reversion to the non-motile phenotype. Alternatively, it could be the result of another process such as angiogenesis (Sec. 6.1.2) through which the tumor gains better access to nutrients. The same model governs the evolution of larger metastatic lesions that have already developed extensive vasculature. Cells die with rate $d$ independent of the number of neighbors, and are replaced by empty sites (non-neoplastic cells within the local tumor environment).

If there is little dispersal ($M \approx 0$), the shape of the tumor becomes roughly spherical as it grows to large size (Fig. 2a, Fig. 6c and Supp. Video 2). However, even a very small amount of dispersal strikingly affects the predicted shape. For $M>0$, the tumor forms a conglomerate of "balls" (Fig. 2b, Fig. 6c, and Supp. Video 3), much like those observed in actual metastatic lesions, with the balls separated by islands of non-neoplastic stromal cells mixed with extracellular matrix. In addition to this remarkable change in topology, dispersal strongly impacts the tumor's growth rate and doubling time. Although the size $N$ of the tumor increases with time $T$ from initiation as $\sim T^3$ without dispersal (Fig. 7), it grows exponentially for $M>0$. This also remains true for long-range dispersal where M plays the role of the reseeding probability of new lesions through circulation from the primary tumor. Using plausible estimates for the rates of cell birth, death, and dispersal probability, we calculate that it takes 8 years for a lesion to grow from one cell to a billion cells in the absence of dispersal ($M=0$), but less than 2 years with dispersal (Fig. 2c). The latter estimate is consistent with experimentally determined rates of metastasis growth as well as clinical experience, while the convention-

al model (without dispersal) is not. In sum, even when individual cancer cells divide and die at identical rates, short-range cell dispersal has a pronounced effect on the size and shape of the resulting tumor.

## 3. Treatment and the evolution of resistance

Non-spatial models point to the size of a tumor as a critical determinant of chemotherapeutic drug resistance [26,27,28,29,30]. To determine whether a spatial model would similarly predict this dependency in a clinically relevant time frame, we calculated tumor regrowth probabilities following targeted therapies. We assume that the cell that initiates the lesion is susceptible to treatment, otherwise the treatment would have no effect on the mass, and that the probability of a resistant mutation is $10^{-7}$ (Sec. 6.1.1); only one such mutation, among all the cells of the metastatic tumor, is needed for a regrowth.

Figure 3a shows snapshots from a simulation (Supp. Video 1) performed before and after the administration of a typical targeted therapy. In this simulation, the metastatic lesion grows from its initiating cell to form a small lesion (~3 mm) when therapy is administered at time $T$=0. The size of the lesion then rapidly decreases, but one month later resistant clones begin to proliferate and form tumors of microscopic size. Such resistant subclones are predicted to be nearly always present in lesions of sizes that can be visualized by clinical imaging techniques [29,30,31]. By six months after treatment, the lesions have regrown to their original size. The evolution of resistance is a stochastic process - some lesions shrink to zero and some regrow (Fig. 8). Figure 3b,c shows the probability of regrowth versus the time from the lesion's initiation to the onset of treatment upon varying net growth rates $b$-$d$ and dispersal probabilities. Regardless of growth rate, the capacity to migrate makes it somewhat more likely that regrowth will occur at shorter periods (compare green and blue curves with red curves in Fig. 3b), particularly for more aggressive cancers, i.e., those which have higher net growth rates (compare Figs. 3b and 3c). This is in line with recent theoretical work on evolving populations of migrating cells [32,33]. If resistant mutations additionally increase the dispersal probability before or during treatment, regrowth is faster (Fig. 8b,c). Note that our model assumes the drug is uniformly distributed in the tumor [34]; it is known that drug gradients can speed up the onset of resistance [35].

## 4. Primary tumors

Having shown that the spatial model's predictions are consistent with metastatic lesion growth and regrowth times, we turn to primary tumors. Here the situation is considerably more complex because the tumor cells are continually acquiring new driver gene mutations that can endow them with fitness advantages over adjacent cells within the same tumor. In contrast, in metastatic lesions we assumed all cells had the same fitness in the absence of targeted therapy. Our model of a primary tumor assumes that it is initiated via a single driver gene mutation that provides a selective growth advantage over normal neighboring cells. Each subsequent driver gene mutation reduces the death rate as $d$=$b(1$-$s)^k$, where $k$ is the number of driver mutations in the cell ($k{\geq}1$), and $s$ is the average fitness advantage per driver. Identical results are obtained if driver gene mutations increase cell birth rather than decrease cell death, or affect both cell birth and cell death (Figs. 9, 11); the only important parameter is the fitness gain, $s$, conferred by each driver mutation.

Figure 4a shows that, in the absence of any new driver mutations (as for a perfectly normal cell growing in utero), clonal subpopulations would be distributed in a conical fashion. Each of these

cones has at least one new genetic alteration, but none of them confers a fitness advantage (they are "passengers"). In an early tumor, in which the center cell harbors the initiating driver gene mutation, the same conical structure would be observed - as long as no new driver gene mutations have yet appeared. The occurrence of a new driver gene mutation, however, dramatically alters the spatial distribution of cells. In particular, the heterogeneity observed in normal cells (Fig. 4a) is substantially reduced (Fig. 4b and Supp. Video 5). The degree of heterogeneity can be quantified by calculating the fraction of genetic alterations (GAs, passengers plus drivers) shared between two cells separated by various distances (Fig. 4d-f). The homogeneity is markedly increased (compare blue vs. red curves in Fig. 4e), even with relatively small fitness advantages (s=1%). This also has implications for the number of GAs that will be present in a macroscopic fraction (e.g., >50%) of all cells. Figure 4f shows that this number is many times larger for s=1% than s=0%. Furthermore, our model predicts that virtually all cells within a large tumor will share at least one new driver gene mutation after 5 years of growth (Fig. 11). The faster the clonal expansion occurs (the larger $s$ is), the smaller the number of passenger GAs (Fig. 11d). Our results are also robust to changes to the model (Sec. 6.4 and Figs. 11, 12). We stress that an important prerequisite for homogeneity is cell turnover in the tumor, because in the spatial setting cells with driver mutations can "percolate" through the tumor only if they replace other cells. In the absence of cell turnover, tumors are much more heterogeneous (Fig. 12d).

## 5. Conclusion

Our model for the spatial evolution of cancers accounts for many facts observed clinically and experimentally. Our results are robust and many assumptions (replication rate proportional to the number of surrounding normal cells, drivers lowering the death rate, constant dispersal rate) can be relaxed without qualitatively affecting the outcome (Sec. 6.4 and SI). Though tumor cell migration has historically been viewed as a feature of cancer associated with late events in tumorigenesis, such as invasion through basement membranes or vascular walls, this classical view of migration pertains to the ability of cancer cells to migrate over large distances [36]. Instead, we focus on small amounts of cellular movement and show that they are able to dramatically reshape a tumor. Moreover, we predict that the rate of tumor growth can be substantially altered by a change in dispersal rate of the cancer cells, even in the absence of any changes in doubling times or net growth rates of the cells within the tumor. Some of our predictions could be experimentally tested using new cell labelling techniques [37,38]. Virtually all previous explanations for the effects of driver gene mutations have assumed that the difference between cell birth and cell death are the major, if not sole, determinants, of cancer growth. Our results suggest that small differences in cell migration can have equally powerful effects on cancer growth - not on individual cells, but on masses of cells in the three-dimensional patterns that actually occur in vivo. This idea could greatly inform the interpretation of mutations in genes whose main functions seem to be related to the cytoskeleton or to cell adhesion rather than to cell birth, death, or differentiation [39,40,41]. For example, cells that have lost the expression of e-cadherin (a cell adhesion protein) are more migratory than normal cells with intact e-cadherin expression [42], and loss of e-cadherin in pancreatic cancer has been associated with poorer prognosis [43], in line with our predictions.

## Acknowledgments



## 6. Methods

### 6.1. Spatial model for tumor evolution.

Tumor modelling has a long tradition [44]. Many models of spatially expanding tumours were proposed in the past [13,14,15,16,17,18,21,45,46,47,48], but they either assume very few [13,15,45,47,49,50,51] or no new mutations at all [14,19,20,46,52,53], or one- or two-dimensional growth [14,15,16,17,54,55]. On the other had, well-mixed models with multiple mutations [6,8,56,57] do not often include space, and more biologically realistic computational models [19,24,58] require too much computing resources (time and memory) to simulate tumors of clinically relevant size (N~$10^9$ cells). Our model builds upon the Eden lattice model [59] and combines spatial growth and accumulation of multiple mutations. Since we focus on the interplay of genetics, spatial expansion, and short-range dispersal of cells, for simplicity we do not explicitly model metabolism [18], tissue mechanics, spatial heterogeneity of tissues, different types of cells present, or angiogenesis [21].

A tumor is made of non-overlapping balls (microlesions) of cells. Tumor cells occupy sites of a regular 3d square lattice (Moore neighborhood, 26 neighbors). Empty lattice sites are assumed to be either normal cells or filled with extracellular matrix and are not modeled explicitly. Each cell in the model is described by its position and a list of genetic alterations (GAs) that have occurred since the initial neoplastic cell, and the information whether a given mutation is a passenger, driver, or resistance-carrying mutation. A passenger mutation does not affect the net growth rate whereas a driver mutation increases it by disrupting tight regulation of cellular divisions and shifts the balance towards increased proliferation or decreased apoptosis. The changes can also be epigenetic and we do not distinguish between different types of alterations. We assume that each GA occurs only once ("infinite allele model" [60]). The average numbers of all GAs, driver, and resistant GAs produced in a single replication event are denoted respetively by $\gamma$, $\gamma_d$, and $\gamma_r$. When a cell replicates, each of the daughter cells receives $n$ new GAs of each type ($n$ being generally different in both cells) drawn at random from the Poisson probability distribution

$$P(n) = \frac{e^{-\gamma_x/2}(\gamma_x/2)^n}{n!}, \tag{1}$$

where "x" denotes the type of GA.

In Model A shown in Figs. 2-4, replication occurs stochastically with rate proportional to the number of empty sites surrounding the replicating cell, and death occurs with constant rate depending only on the number of drivers. We also simulated other scenarios (Models B, C, D, see below). Driver mutations increase the net growth rate (the difference between proliferation and death) either by increasing the birth rate or decreasing the death rate by a constant factor $1+s$ where $s>0$.

Dispersal is modelled by moving an offspring cell to a nearby position where it starts a new microlesion (Fig. 5a). Microlesions repel each other; a "shoving" algorithm [61,62] (Fig. 5b) ensures they do not merge.

The computer code can handle up to $10^9$ cells, which corresponds to tumors that are clinically meaningful and can be observed by conventional medical imaging (diameter >1cm). The algorithm is discussed in details in the Supplementary Information. It is not an exact Kinetic Monte Carlo (KMC) algorithm because such an algorithm would be too slow to simulate large tumors. A comparison with KMC for smaller tumors (Supplementary Information and Fig. 14) shows that both algorithms produce consistent results.

### 6.1.1. Model parameters

The initial birth rate $b$=ln $2 \approx 0.69$ days$^{-1}$, which corresponds to a 24h minimum doubling time. The initial death rate $d$=0...0.995$b$ depends on the aggressiveness of the tumor (larger values = less aggressive lesion). In simulations of targeted therapy, we assume that, prior to treatment, $b$=0.69 days$^{-1}$ and $d$=0.5$b$=0.35 days$^{-1}$, whereas during treatment $b$=0.35 days$^{-1}$ and $d$=0.69 days$^{-1}$, i.e. birth and death rates swap places. This rather arbitrary choice leads to the regrowth time of about 6 months which agrees well with clinical evidence. Mutation probabilities are $\gamma$=0.02, $\gamma_d$=4·10$^{-5}$, $\gamma_r$=10$^{-7}$, in line with experimental evidence and theoretical work [8,63,64,65]. Since there are no reliable data on the dispersal probability $M$, we have explored a range of values between $M$=10$^{-7}$ and 10$^{-2}$. All parameters are summarized in Table 2, see also further discussion in Supplementary Information.

### 6.1.2. Validity of the assumptions of the model

Our model is deliberately oversimplified. However, many of the assumptions we make can be experimentally justified or shown not to qualitatively affect the model.

**Three-dimensional regular lattice of cells.** The 3d Moore neighborhood was chosen because it is computationally fast and introduces relatively fewer artifacts related to lattice symmetries. Real tissues are much less regular and the number of nearest neighbors is different [66]. However, recent simulations of similar models of bacterial colonies [67,68] show that the structure of the lattice (or the lack thereof in off-lattice models) has a marginal effect on genetic heterogeneity.

**Asynchronous cell division.** Division times of related cells remain correlated for a few generations. However, stochastic cell division implemented in our model is a good approximation for a large mass of cells and is much less computationally expensive than modelling a full cell cycle.

**Replication faster at the boundary than in the interior.** Several studies have described a higher proliferation rate at the leading edge of tumors, and this has been associated with a more aggressive clinical course [69]. To estimate the range of death rate $d$ for our model, we used the proliferation marker Ki-67. Representative formalin-fixed, paraffin-embedded tissue blocks were selected from 4 small chromophobe renal cell carcinomas and 6 small hepatocellular carcinomas by the pathologist (M.E.P). A section of each block was immunolabeled for Ki67 using the Ventana Benchmark XT system. Eight to 12 images, depending on the size of the lesion, were acquired from each tumor. Fields were chosen at random from the leading edge and the middle of the tumor and were not necessarily "hot spots" of proliferative activity. Using an ImageJ macro, each Ki-67 positive tumor nucleus was labeled red by the pathologist, and each Ki67 negative tumor nucleus was labeled green. Other cell types (endothelium, fibroblasts, inflammatory cells) were not labeled. The proliferation rate was then calculated using previously described methods [70]. The study was approved by the Institutional Review Board of the Johns Hopkins University School of Medicine. In all ten tumors the proliferation rate at the leading edge of the tumor was greater than that at the center by a factor of 1.25 to 6 (Table 1). Comparing the density of proliferating cells to our model gives $d \approx 0.5b$ (range: $d$=0.17$b$...0.8$b$), which is what we assume in the simulations of aggressive lesions.

**Equal fitness of all cells in metastatic lesions.** We assume that cells in a metastatic lesion are already very fit since they contain multiple drivers. Experimental evidence in microbes [71] and (to a lesser extent) in eukaryotes [72] suggests that fitness gains due to individual mutations are largest at the beginning of an evolutionary process and that the effects of later mutations are much smaller. Consequently, new drivers occurring in the lesion will not be able to spread through the population before the lesion reaches a clinically relevant size. Indeed, studies of primary tumors and their

matched metastases usually fail to find driver mutations present in the metastases that were not present in the primary lesions [2,73].

**Dispersal.** In our model, cells detach from the lesion and attach again at a different location in the tissue. This can be viewed either as cells migrating from one place to another one, or as a more generic mechanism that allows tumor cells to get better access to nutrients by dispersing within the tissue, hence providing a growth advantage over cells that did not disperse. Some mechanisms that do not involve active motion (i.e. cells becoming motile) are discussed below.

**Migration.** Cancer cells are known to undergo epithelial-to-mesenchymal transition (EMT), whose origin is thought to be epigenetic [25]. This involves a cell becoming motile and moving some distance. If the cell finds the right environment, it can switch back to the non-motile phenotype and start a new lesion. Motility can be enhanced by tissue fluidization due to replication and death [74]. Instead of modelling the entire cycle (epithelial-mesenchymal-epithelial), we only model the final outcome (a cell has moved some distance).

**Tumor buds.** Many tumors exhibit focally invasive cell clusters, also known as tumor buds. Their proliferation rate is less than that of cells in the main tumor [75]. We hypothesize that tumor buds contain cells that have not yet completed EMT and therefore they proliferate slower.

**Single versus cluster migration.** Ref. [76] found that circulating cancer cells (CRCs) can travel in clusters of 2-50 cells and that such CRC clusters can initiate metastatic foci. They report that approximately half of the metastatic foci they examined were initiated by single CRCs, and that CRC clusters initiated the other half. The authors also note that the cells forming a cluster are most likely neighboring cancer cells from the primary tumor. This means that the genetic makeup of cells within a newly established lesion will be very similar, regardless of its origin (single cell versus a small cluster of cells). Therefore the ability to travel in clusters should not affect the genetic heterogeneity or regrowth probability as compared to single-cell dispersal from our model.

**Angiogenesis.** We do not explicitly model angiogenesis for two reasons. First, most GAs that can either change the growth rate or be detected experimentally must occur at early stages of tumor growth as explained before. Hence, the genetic make-up of the tumor is determined primarily by what happens before angiogenesis. Second, local dispersal from the model mimics tumor cells interspersing with the vascularized tissue and getting better access to nutrients, which is one of the outcomes of angiogenesis.

**Biomechanics of tumors.** Growth is affected by the mechanical properties of cells and the extracellular matrix. We do not explicitly include biomechanics, in contrast to more realistic models [77,78], as this would not allow us to simulate lesions larger than about $10^6$ cells. Instead, we take experimentally determined values for birth and death rates, values that are affected by biomechanics, as the parameters of our model.

**Isolated balls of cells.** In our simulations, balls of cells are thought to be separated by normal, vascularized tissue which delivers nutrients to the tumor. Each ball's environment is the same and there are no interactions between the balls other than mechanical repulsion. This represents a convenient mathematical contrivance and qualitatively recapitulates what is observed in stained sections of actual tumors (Fig. 1a). However, in reality microlesions within the primary tumor are neither spherical nor completely separated, and some of them are better described as "protrusions" bulging out from the main tumor tissue due to biomechanical instabilities [79]. Nevertheless, we believe that modelling the tumor as a collection of non- or weakly-interacting microlesions is essentially correct. Many tumor cells are less adhesive and more elastic than normal cells [80], and it is known that dif-

ferences in cellular adhesion and stiffness promote segregation of different types of cells [81,82]. Thus, if a cancer cell migrates far enough from the surface of the lesion, its progeny will form a cluster of cells rather than mix with the normal tissue. For the same reason, and unlike in the model, nearby microlesions may partially merge. However, the capillary network of blood vessels - either due to tumor angiogenesis or preexisting in the invaded tissue - may still provide enough nutrients to the lesions so that our assumption of independently growing balls of cells remains valid.

## 6.2. Tumor geometry and heterogeneity in the absence of driver mutations

Supplementary videos 2 and 3 illustrate the process of growth of a tumor with maximally $N=10^7$ cells, for $M=0$ and $M=10^{-6}$, respectively, and for $d=0.5$. Figure 6 shows snapshots from a single simulation for $M=0$, $N\sim10^3$, and $d=0$ (no death, panel a) and $d=0.9$ (panel b). In the latter case, cells are separated by empty sites (normal cells/extracellular matrix). Panel (c) shows that the tumor is almost spherically symmetric for $M=0$. The symmetry is lost for small but non-zero $M$, and restored for larger $M$ when the balls become smaller and their number increases. Panel (c) also shows that metastatic tumors contain many clonal sectors with passenger mutations. Figure 10a shows that the fraction $G(r)$ of GAs that are the same in two randomly sampled cells (cf. Fig. 4) separated by distance $r$ quickly decreases with $r$, indicating increased genetic heterogeneity due to passenger mutations.

## 6.3. Targeted therapy of metastatic lesions

Models of cancer treatment [29,30,34,83,84,85] often assume either no spatial structure or do not model the emergence of resistance. We assume that the cell that initiated the lesion was sensitive to treatment but its progeny may become resistant. Before the therapy commences, all cells have the same birth and death rates, but after the treatment resistant cells continue to proliferate with the same rate whereas susceptible cells are assigned different rates as described above. Resistant cells can emerge prior to and during the therapy.

Supplementary Video 1 and Figure 8a show that, since the process of resistance acquisition is stochastic, some tumors regrow after an initial regression, and some do not. If only resistant cells can migrate, regrowth is faster (Fig. 8b,c). Figure 8d-g shows regrowth probabilities $P_{regrowth}$ for different treatment scenarios not mentioned in the main text, depending on whether the drug is cytostatic ($b_{treatment}=0$) or cytocidal ($d_{treatment} = b$), and whether $d=0$ or $d>0$ before treatment. In Panel (d), cells replicate and die only on the surface, and the core is "quiescent" - cells are still alive there but cannot replicate unless outer layers are removed by treatment (Supp. Video 6 and 7). $P_{regrowth}$ does not depend on the migration rate $M$ at all, and is close to 100% for $N>10^8$ cells, a size that is larger than for $d>0$ (Panel (f)). It can be shown that $P_{regrowth} = 1-\exp(-\gamma_r N)$. Panel (e) is for the cytostatic drug ($b_{treatment}=d_{treatment}=0$); this is also equivalent to the cytocidal drug if the tumor has a necrotic core (cells are dead but still occupy physical volume). In this case, $P_{regrowth}$ increases with $M$ because more resistant cells are on the surface for larger $M$ (cells can replicate only on the surface in this scenario). Figure 8f,g shows models with cell death present even in the absence of treatment ($d=0.9b$) but occurring only at the surface, unlike in Fig. 3 where cells also die inside the tumor. Death increases $P_{regrowth}$ due to a larger number of cellular division necessary to obtain the same size, and hence more opportunities to mutate.

## 6.4. Relaxing the assumptions of the model

Figure 4 shows that even a small fitness advantage substantially reduces genetic diversity through the process of clonal expansion, see also Suppl. Videos 4 and 5. We now demonstrate that this also applies to modified versions of the model, proving its robustness.

**Only the net growth rate matters.** Figure 9 shows that an advantageous mutation with $s=1\%$ spreads through the tumor regardless of whether drivers affect death (upper panels) or growth (lower panels), also in the presence of non-zero migration. This is further confirmed in Fig. 10b,e, which shows that the average number of shared GAs is larger in the presence of drivers. Figure 10c,f shows that as long as $s>0$ and regardless of its exact value, driver mutations homogenize the tumor compared to the case $s=0$. Figure 11a-c shows how many driver mutations are expected to be present in a randomly chosen cell from a tumor that is $T$ years old. Neither migration nor the way drivers affect growth (via birth or death rate) has significant effect on the number of drivers per cell (Panels b,c). A small discrepancy visible in Panel (b) is caused by a slightly asymmetric way death and birth is treated in our model, see the Supplementary Information.

**Model B**. In this modified model, cells replicate with constant rate if there is at least one empty neighbor. In the absence of drivers, GAs are distributed evenly throughout the lesion (Fig. 12b) but they often occur independently and the number of frequent GAs is low (Fig. 12e). Drivers cause clonal expansion as in Model A.

**Model C**. Cells replicate regardless of whether there are empty sites surrounding them or not. When a cell replicates, it pushes away other cells towards the surface (Supplementary Information). Figure 12c,e shows that this again leads to clonal expansion which decreases diversity.

**Model D**. Replication/death occurs only on the surface and the core of the tumor is static. Figure 12d shows that driver mutations cannot spread to the other side of the lesion and conical clonal sectors can be seen even for $s>0$. The number of frequent GAs is the same for $s=0$ and $s=1\%$ indicating that genetic heterogeneity is not lowered by clonal expansion. This demonstrates that cell turnover inside the tumor is very important for reducing heterogeneity. To obtain the same (low) heterogeneity as for Models A-C, the probability of driver mutations must be much larger in Model D (Fig. 12f).

**Drivers affecting M.** We investigated three scenarios in which drivers affect (1) only the migration rate $M\rightarrow(1+q)M$ where $q>0$ is the "migration fitness advantage" (no change in $b,d$), (2) both $M$ and $d$, i.e. $(d,M)\rightarrow(d(1-s),(1+q)M)$ with $s,q>0$, (3) either $M$ or $d$, with probability 1/2. Figure 13 shows that growth is unaffected in cases (1,3) compared to the neutral case. For (2) the tumor growth rate increases significantly when the tumor is larger than $N=10^6$ cells. This shows that migration increases the overall fitness advantage, in line with Ref. [86] which shows that fixation probability is determined by the product of the exponential growth rate and diffusion constant (motility) of organisms.

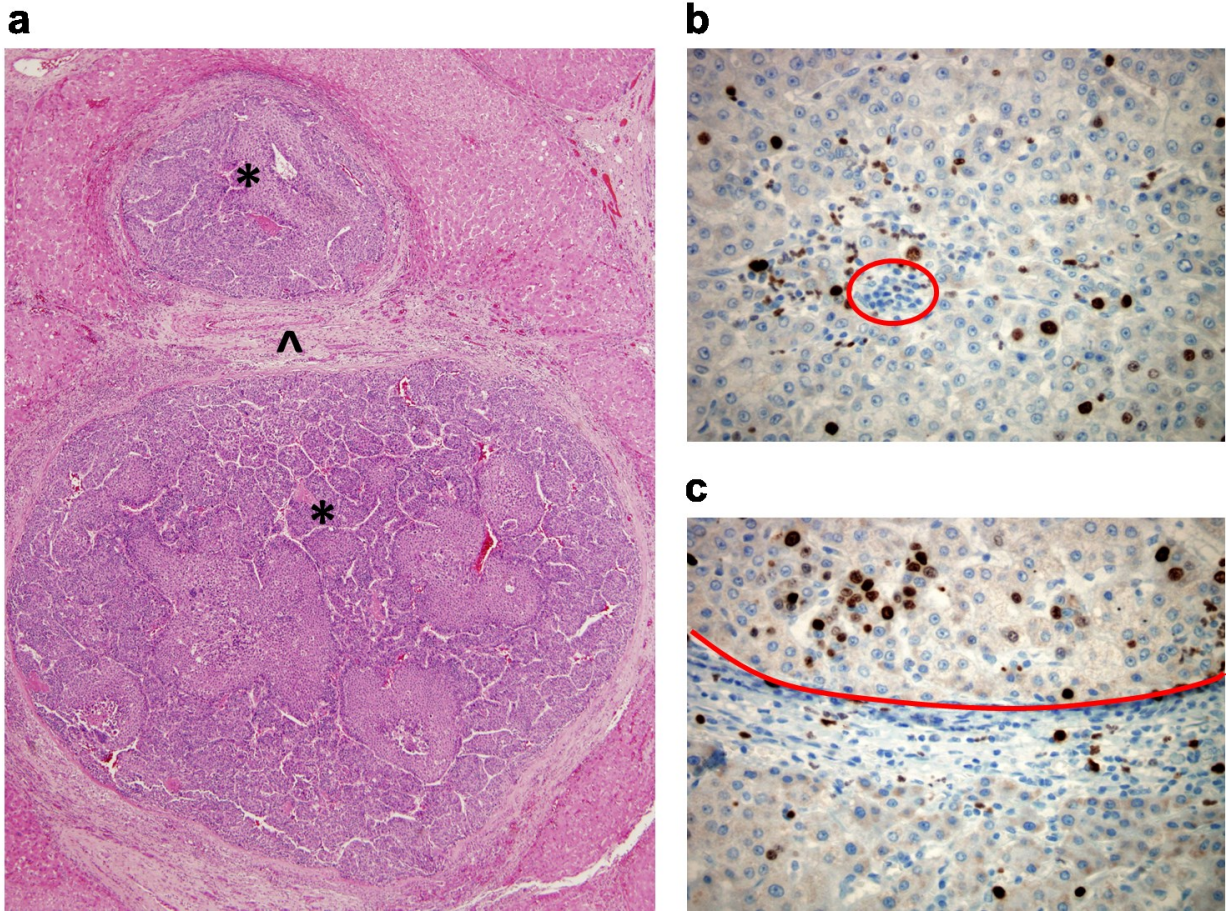

Fig. 1. **Structure of solid tumors**. **(a)** Hematoxylin and eosin stained section of hepatocellular carcinoma. The neoplasm is composed of two distinct balls of cells (*) separated by non-tumor tissue (^). Note the smaller nests (balls) of neoplastic cells within the lower lesion. **(b-c)** Hepatocellular carcinoma immunolabeled with the proliferation marker Ki67. The middle of the tumor (b) shows decreased proliferation as compared to the leading edge of the tumor (top of c), see Table 1. The red circle in (b) demonstrates an example of cells (inflammatory cells) that would not have been included in the total cell count. The areas below the red line in (c) are non-tumor stroma and non-tumor liver tissue.

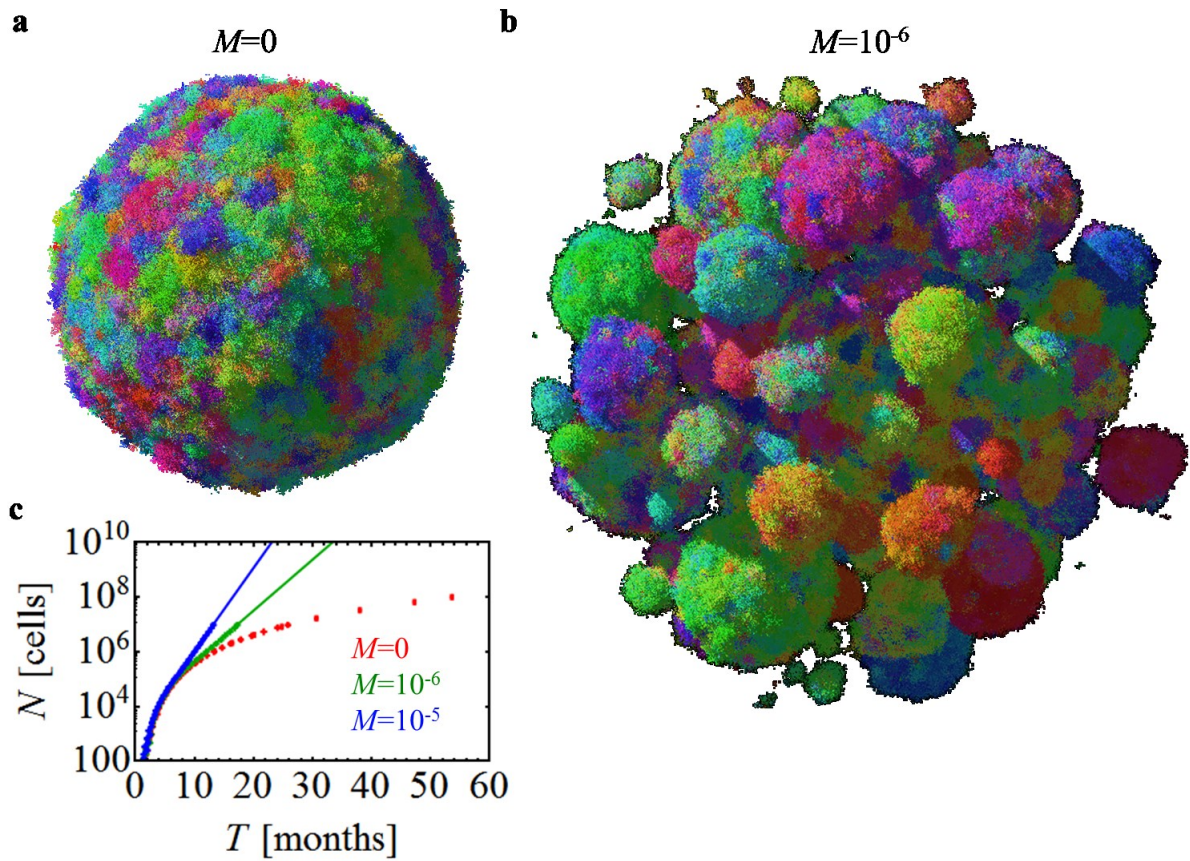

Fig. 2. **Short-range migration affects size, shape, and growth rate of tumors.** (**a**) A spherical lesion in the absence of migration ($M$=0) and (**b**) a conglomerate of lesions, each initiated by a cell that has migrated from a prior lesion, for low but non-zero migration ($M$=$10^{-6}$). Colors reflect the degree of genetic similarity; cells with similar colors have similar genetic alterations (GAs). The death rate is $d$=$0.8b$ which corresponds to a net growth rate of $0.2b$=$0.14$days$^{-1}$, and $N$=$10^7$ cells. (**c**) The number of cancer cells $N$ increases in time much faster for $M$>$0$ than in the absence of migration.

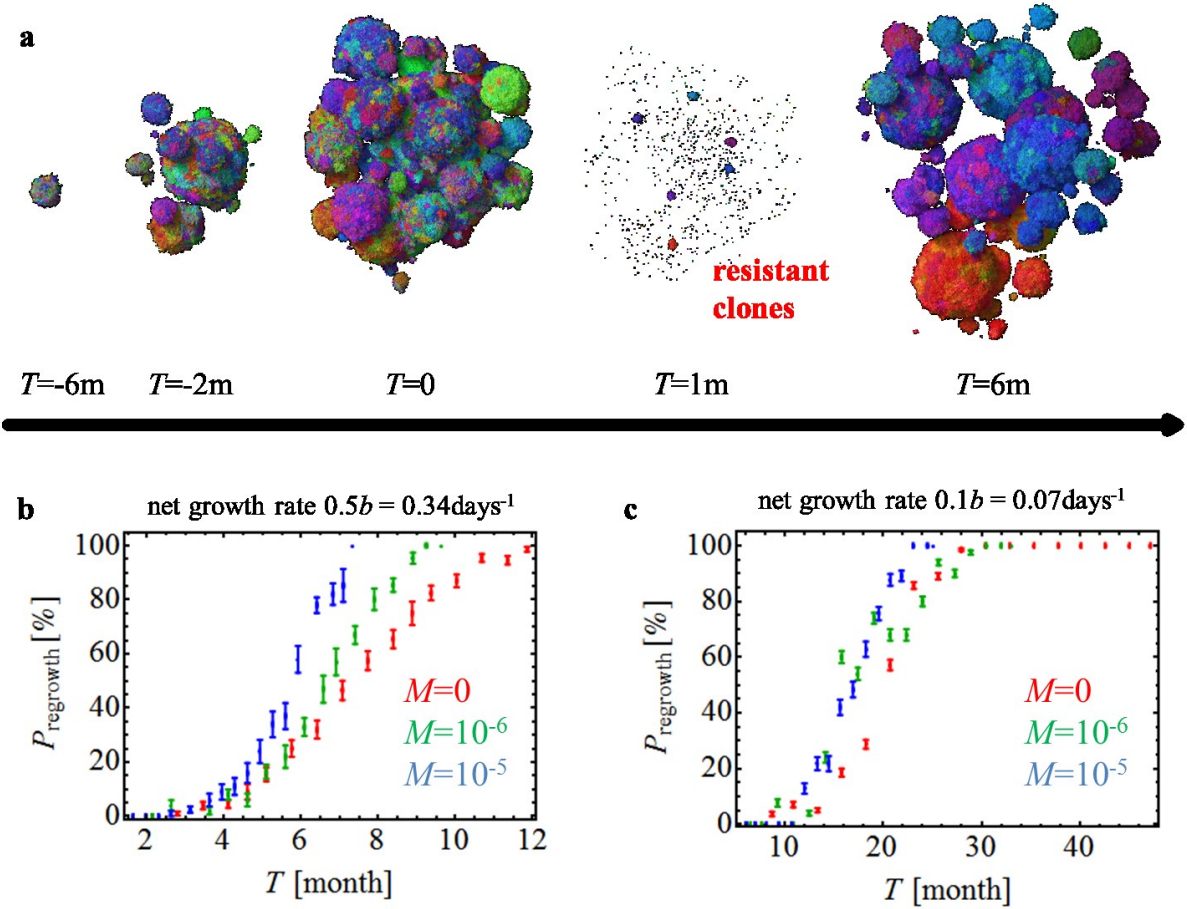

Fig. 3. **Treatment success rates depend on the net growth rate of tumors.** (**a**) Time snapshots prior to and during therapy ("m" = months). Treatment increases the death rate and decreases the growth rate of susceptible cells throughout the tumor. Resistant subpopulations that cause the tumor to regrow after treatment can be seen at $T$=1m. Parameters: $d$=0.5$b$ in the absence of treatment, $M$=10$^{-6}$; treatment begins when the tumor has $N$=10$^7$ cells. (**b-c**) Probability of tumor regrowth ($P_{regrowth}$) as a function of time after treatment initiation, for different migration probabilities ($M$) and net growth rates of the resistant cells. A higher growth rate (**b**) (either because of a lower death rate $d$ or equivalently a higher birth rate $b$) leads to a high regrowth probability, so that 50% of tumors regrow six months after treatment is initiated when $M$=10$^{-5}$. (**c**) Tumors with lower net growth rates (i.e., higher death rates) require >20 months to achieve the same probability of regrowth. In both (b) and (c) we assume that the growth rate of resistant cells after therapy is identical to that of the sensitive cells prior to treatment. The death rate of sensitive cells during treatment is greater than the birth rate, or the tumors would not be sensitive to the drug.

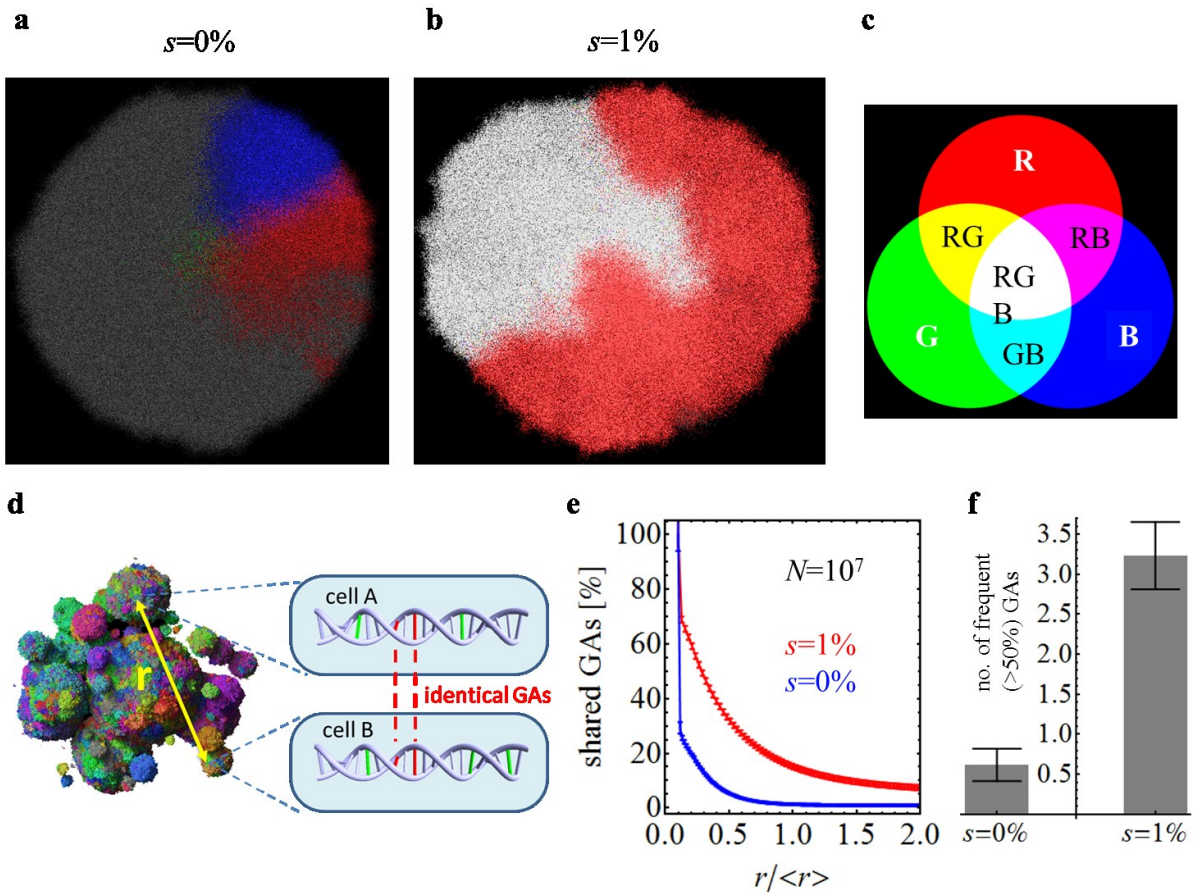

Fig. 4. **Genetic diversity is strongly reduced by the emergence of driver mutations.** Genetic diversity within a single lesion ($M$=0) with initial death rate $d$=0.99$b$ (net growth rate = 0.007days⁻¹). The three most abundant genetic alterations (GAs) have been color-coded using red, green, and blue (see panel **c**). Combinations of the three basic colors correspond to cells having 2 or 3 of these GAs. **(a)** No drivers – separated, conical sectors emerge in different part of the lesion, each corresponding to a different clone. **(b)** Drivers with fitness advantage s=1% lead to clonal expansions and many cells have all 3 GAs (white area). **(d)** Drawing shows how genetic diversity can be determined quantitatively: genotypes of two randomly chosen cells, separated by distance $r$, are compared and the number of shared GAs is determined. **(e)** The number of shared GAs versus the normalized distance $r/<r>$ decreases much slower for the case with (red) than without driver mutations (blue). **(f)** The total number of GAs present in at least 50% of all cells is much larger for s=1% than for s=0%.

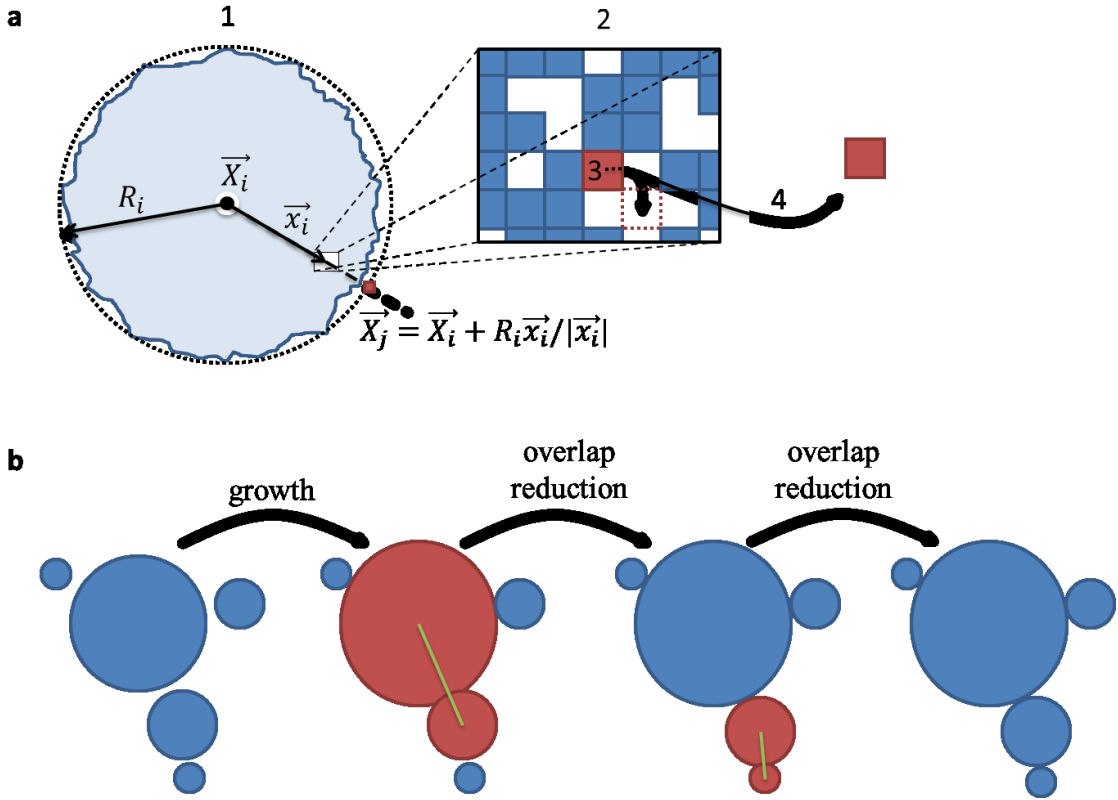

Fig. 5. **Details of the model.** (**a**) A sketch showing how dispersal is implemented: (1) A ball of cells of radius $R_i$ which center is at $X_i$ is composed of tumor cells and normal cells (blue and empty squares in the zoomed-in rectangle (2)). A cell at position $x_i$ w.r.to the center of the ball attempts to replicate (3). If replication is successful, the cell migrates with probability $M$ and creates a new microlesion (4). The position $X_j$ of this new ball of cells is determined as the endpoint of the vector which starts at $X_i$ and has direction $x_i$ and length $R_i$. (**b**) Overlap reduction between the balls of cells. When a growing ball begins to overlap with another ball (red), they are both moved apart along the line connecting their centers of mass (green line) by as much as necessary to reduce the overlap to zero. The process is repeated for all overlapping balls as many times as needed until there is no overlap.

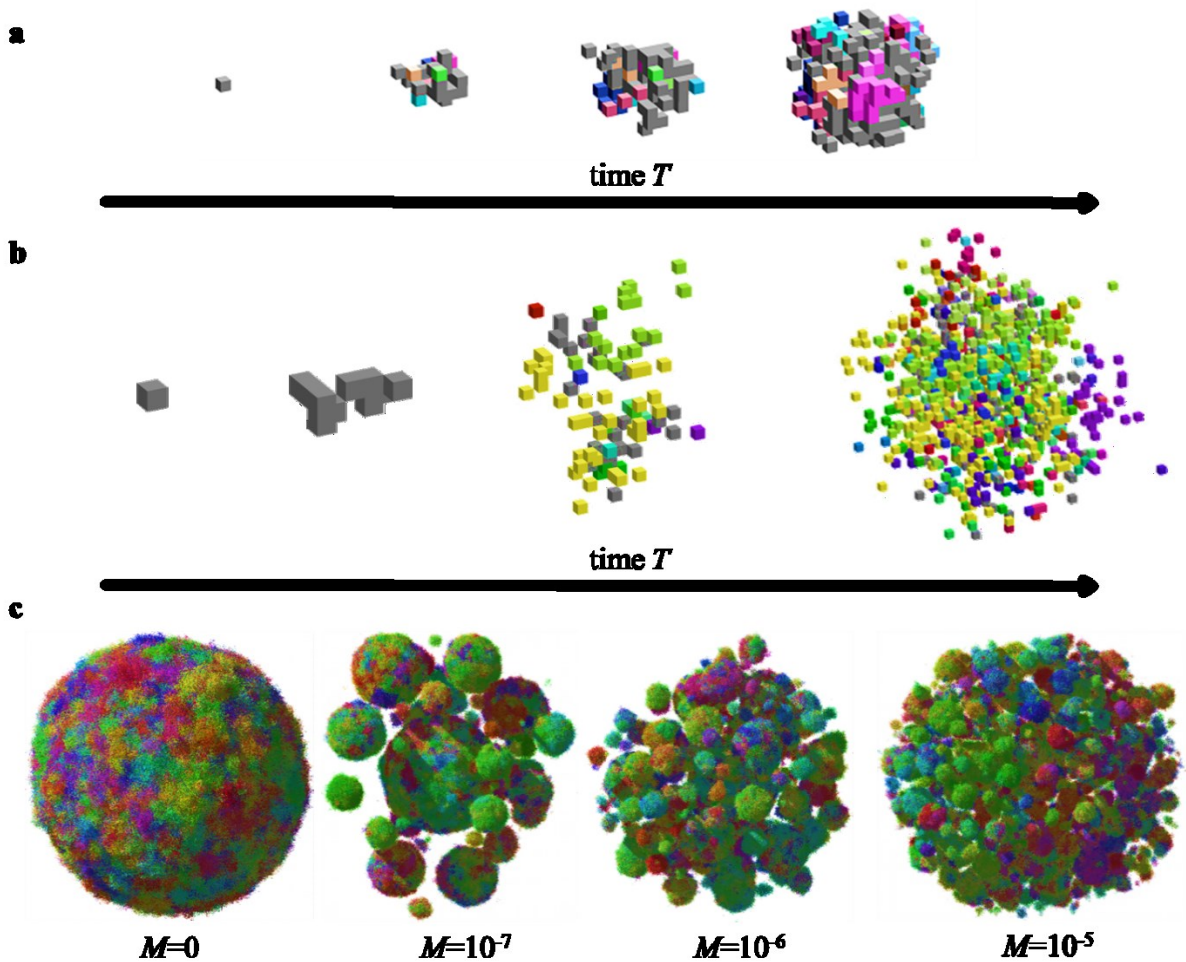

Fig. 6. **Simulation snapshots. (a-b)** A few snapshots of tumor growth for no dispersal and (a) $d$=0 (b) $d$=0.9$b$. To visualize clonal sectors, cells have been color-coded by making the color a heritable trait and changing each of its RGB components by a small random fraction whenever a cell mutates. The initial cell is gray. Empty space (white) are stromal cells mixed with extracellular matrix. Note that images are not to scale. **(c)** Tumor shapes for $N$=10$^7$, $d$=0.9$b$, and different dispersal probability $M$. Images not to scale – the tumor for M=10$^{-5}$ is larger than the one for M=0.

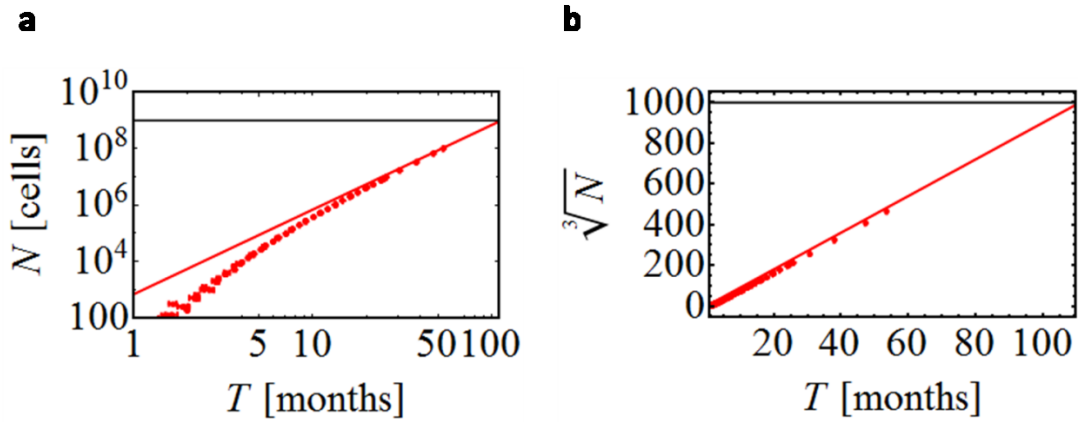

Fig. 7. **Tumor size as a function of time. (a)** Growth of a tumor without dispersal ($M$=0), for $d$=0.8$b$. For large times ($T$), the number of cells grows approximately as const×$T^3$. The tumor reaches size $N$=$10^9$ cells (horizontal line) after about 100 months (8 years) of growth. **(b)** The same data is plotted in the linear scale, with $N$ replaced by "linear extension" $N^{1/3}$.

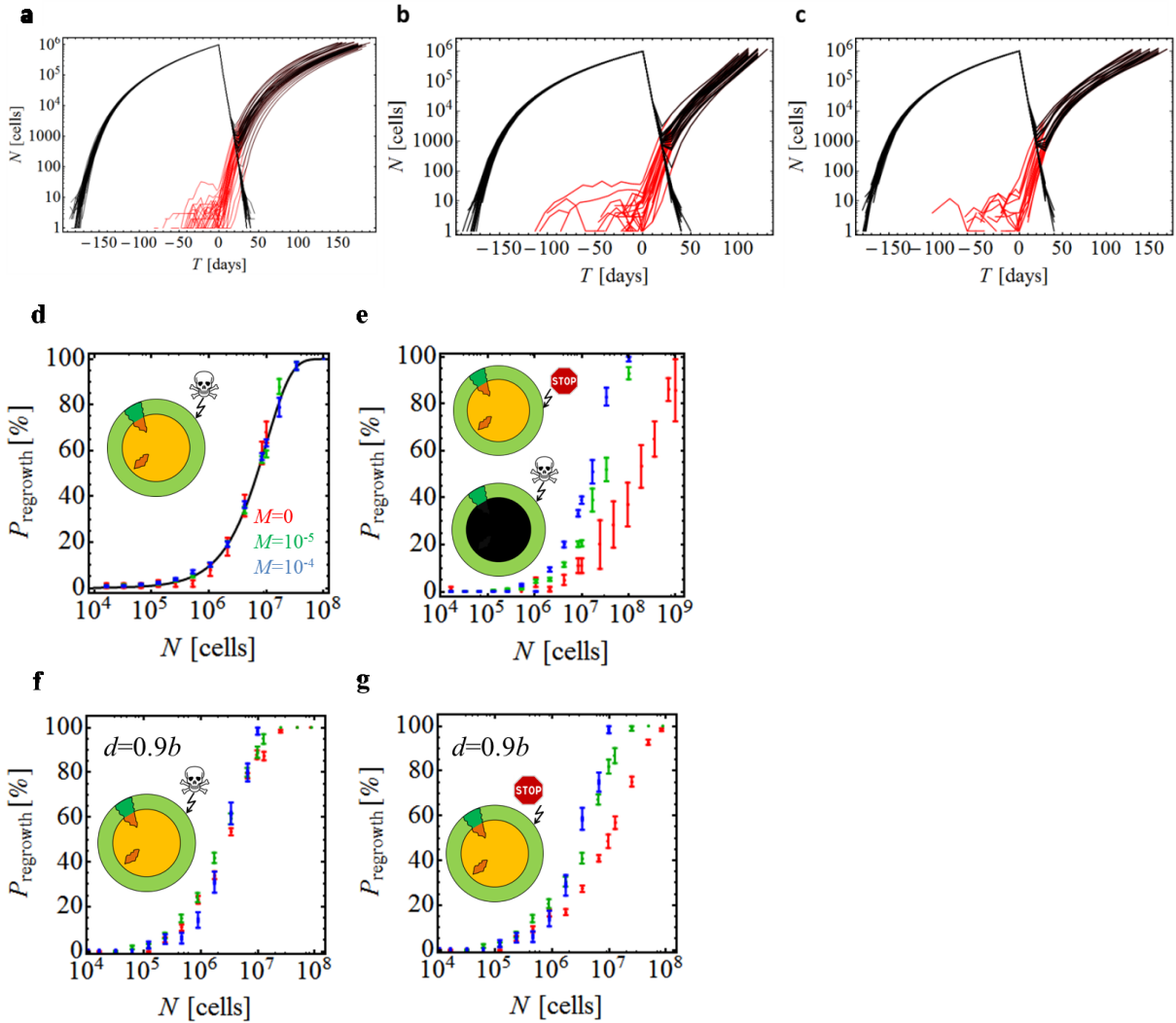

Fig. 8. **Simulation of targeted therapy.** (**a-c**) The total number of cells in the tumor (black) and the number or resistant cells (red) versus time, during growth ($T<0$) and treatment ($T>0$), for ~100 independent simulations, for $d=0.5b$ for $T<0$. Therapy begins when $N=10^6$ cells. Upon treatment, many tumors die out ($N$ decreases to zero) but those with resistant cells will regrow sooner or later. (**a**) $M=0$ for all cells at all times. (**b**) M=0 for all cells for T<0 and M=$10^{-4}$ for resistant cells for T>0. (**c**) M=0 for non-resistant and M=$10^{-5}$ for resistant cells at all times. In all three cases, $P_{regrowth}$ is very similar: 36+/-5% for (a), 25+/-4% for (b), and 27+/-6% for (c). (**d-g**) Regrowth probability for four treatment scenarios not discussed in the main text. Data points correspond to three dispersal probabilities: $M=0$ (red), $M=10^{-5}$ (green), and $M=10^{-4}$ (blue). The probability is plotted as a function of tumor size $N$ just before the therapy commences. (**d**) Before treatment, cells replicate only on the surface. Cells in the core are quiescent and do not replicate. Therapy kills cells on the surface and in the core resume proliferation when liberated by treatment. (**e**) As in (d) but drug is cytostatic and does not kill cells but inhibits their growth. The results for $P_{regrowth}$ are identical if the drug is cytotoxic and the tumor has a necrotic core (cells die inside the tumor and cannot replicate even if the surface is removed). (**f**) Before treatment, cells replicate and die on the surface. The core is quiescent. Therapy kills cells on the surface (cytotoxic drug). (**g**) As in (f) but therapy only inhibits growth (cytostatic drug).

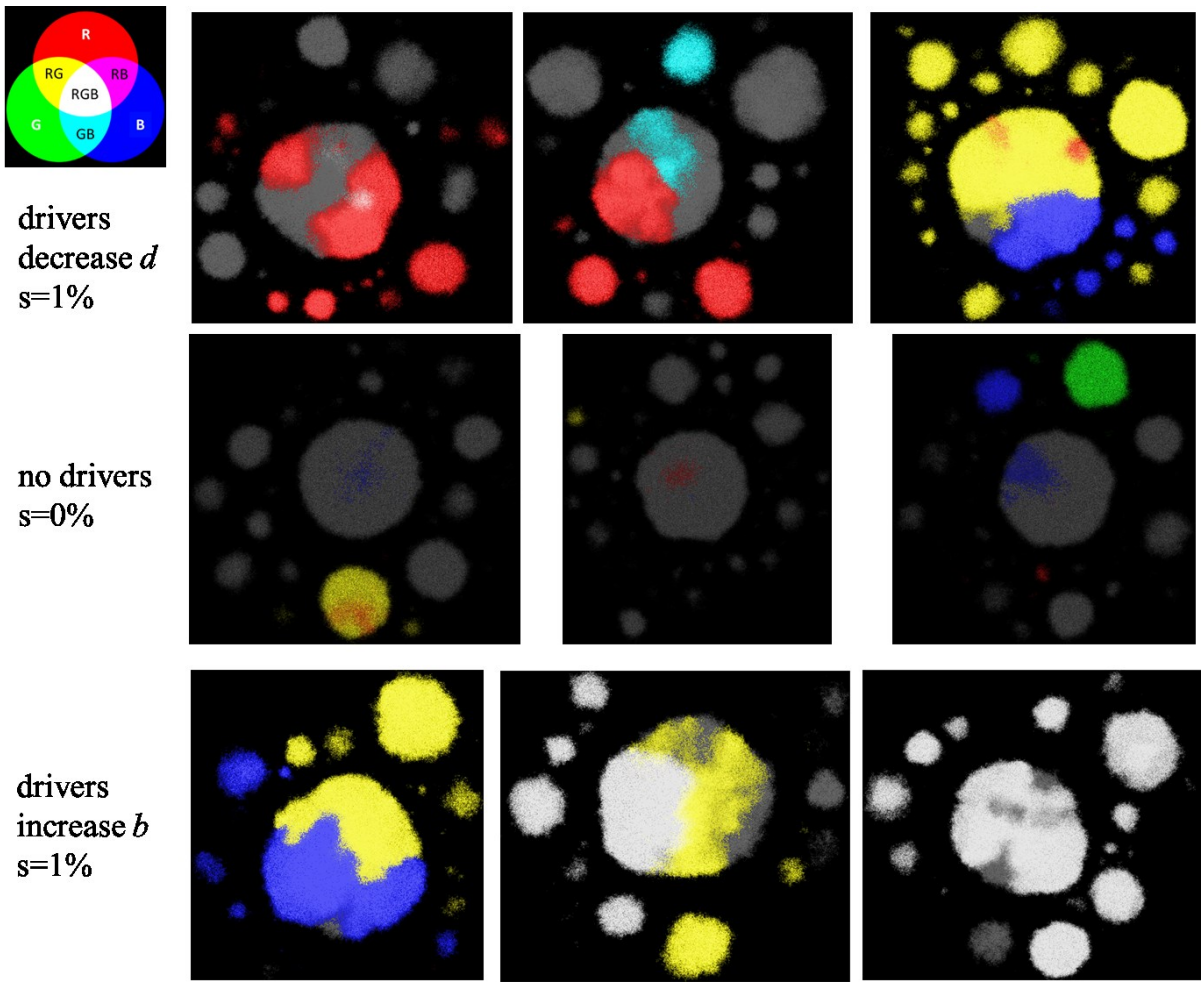

Fig. 9. **Most abundant GAs in the presence of drivers and dispersal.** Sections through the center of the tumor (layer of thickness of 80 cells and symmetric about the center) for $N$=10$^7$, dispersal probability $M$=10$^{-7}$, with three most abundant mutations color-coded as red, green, and blue (as in Fig. 4). Three columns correspond to three representative independent simulation runs. A small color-mixing palette in the top-left corner explains how these three mutations combine to produce other colors. Upper panel: drivers decrease the rate of death (as in the main text). Lower panel: drivers increase the rate of growth. Middle panel: no driver mutations (neutral case with $s$=0% and $d$=0.99$b$).

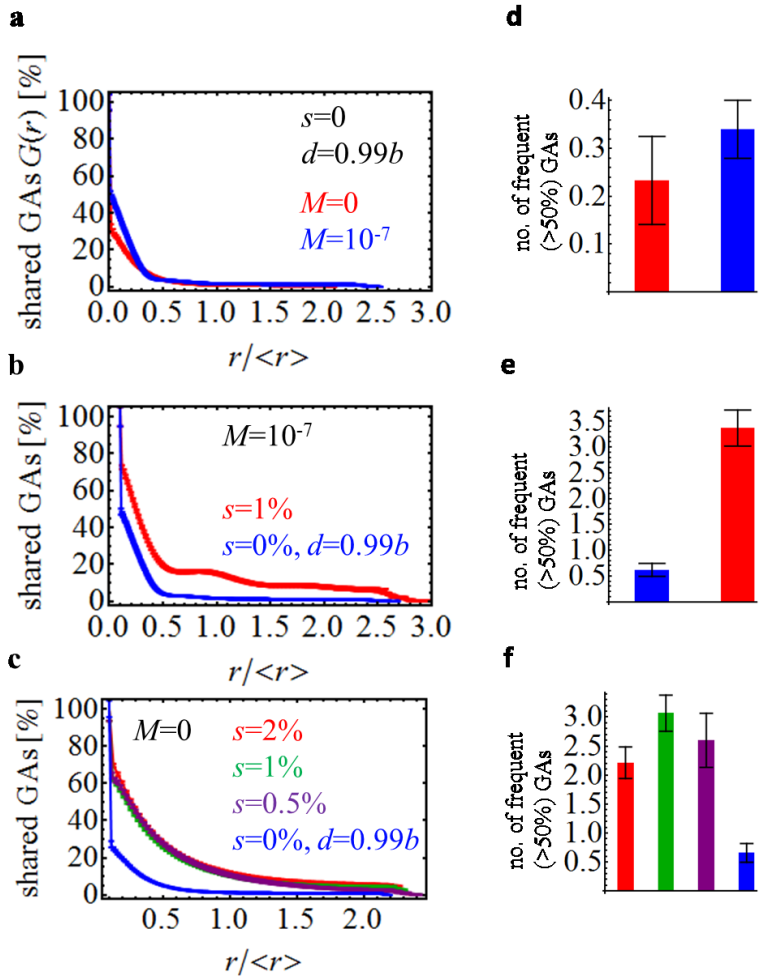

Fig. 10. **Genetic diversity quantified.** **(a)** Tumors are much more genetically heterogeneous in the absence of driver mutations ($s$=0). The plot shows the fraction $G(r)$ of genetic alterations (GAs) shared between the cells as function of their separation (distance $r$) in the tumor. The fraction quickly decreases with increasing $r$. The distance in the figure is normalized by the average distance $<r>$ between any two cells in the tumor. For a spherical tumor, $<r>$ is approximately equal to half of the tumor diameter. **(b)** Fraction of shared GAs for $s$=1% and $s$=0%, $N$=$10^7$, and $M$=$10^{-7}$. In the presence of drivers, $G(r)$ decays slower, indicating more homogeneous tumors. **(c)** The exact value of the selective advantage of driver mutations is not important (all curves GAs($r$) look the same, except for $s$=0) as long as $s$>0. **(d,e,f)**: Number of GAs present in at least 50% of cells for identical parameters as in panels (a,b,c), correspondingly. Drivers substantially increase the level of genetic homogeneity.

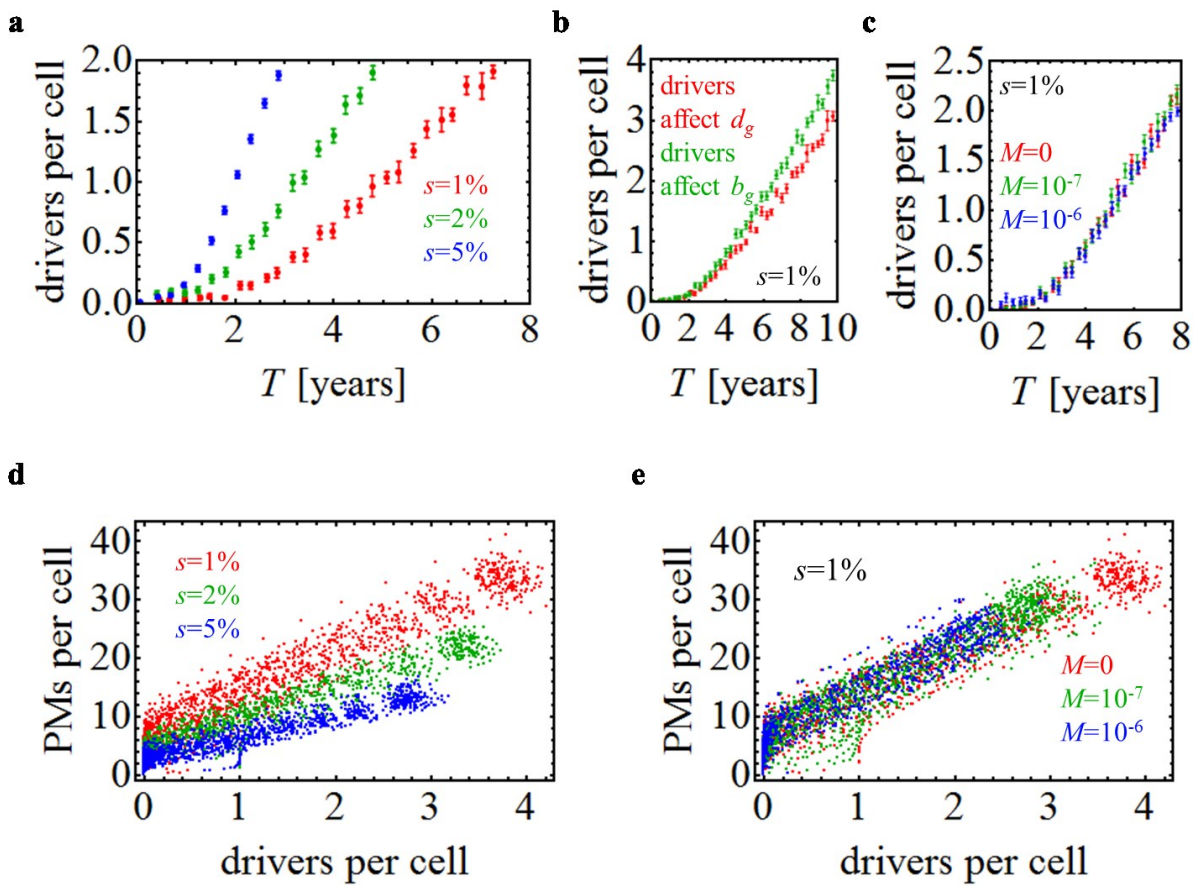

Fig. 11. **Accumulation of driver and passenger GAs. (a-c)** The number of drivers per cell in the primary tumor plotted as a function of time. **(a)** $M=0$ and three different driver selective advantages. For $s=1\%$, cells accumulate on average one driver mutation within 5 years. The time can be significantly lower for very strong drivers ($s>1\%$). **(b)** The rate at which drivers accumulate depends mainly on their selective advantage and not on whether they affect death or birth rate. **(c)** Dispersal does not affect the rate of driver accumulation. **(d-e)** The number of passenger mutations (PMs) per cell versus the number of driver mutations per cell. More PMs are present for smaller driver selective advantage (panel d) and this is independent of the dispersal probability $M$ (panel e) in the regime of small $M$. Data points correspond to independent simulations.

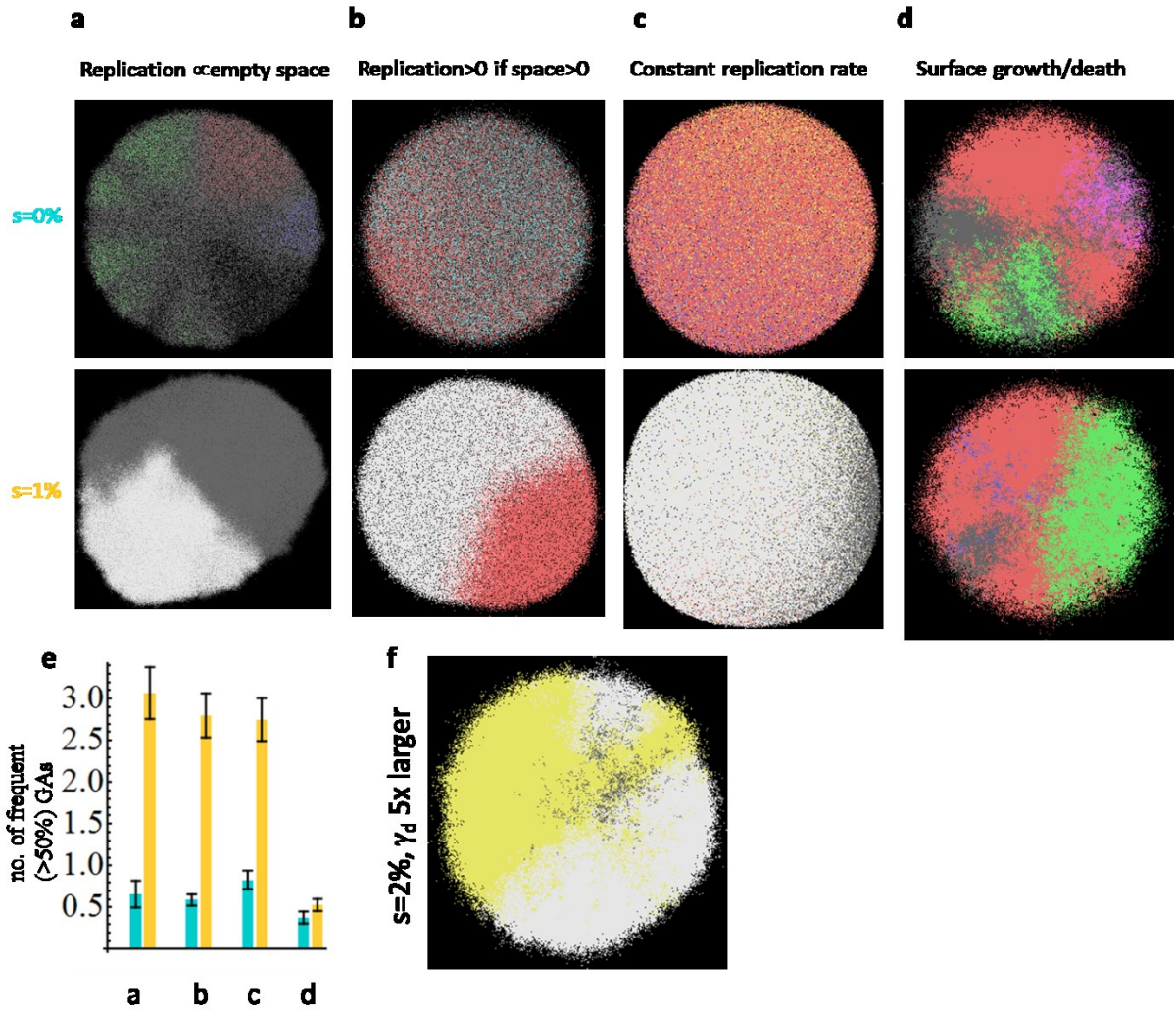

Fig. 12. **Genetic diversity in a single lesion for different models. (a-d)** Representative simulation snapshots, with GAs color-coded as in Fig. 4. Upper panels: $s$=0, lower panels: $s$=1%. Panel (**a**): Model A from the main text in which cells replicate with rates proportional to the number of empty nearby sites. Panel (**b**): Model B – the replication rate is constant and non-zero if there is at least one empty site nearby, and zero otherwise. Panel (**c**): Model C - cells replicate at a constant rate and push away other cells to make space for their progeny. Panel (**d**): Model D - cells replicate/die only on the surface, the interior of the tumor ("necrotic core") is static. In all cases $N$=$10^7$, $d$=0.99$b$. (**e**) Number of GAs present in at least 50% of cells for identical parameters as in panels a-d. In all cases except surface growth (d), drivers increase genetic homogeneity, measured by the number of frequent GAs. (**f**) Model D with $\gamma_d$ =2x$10^{-4}$ instead 4x$10^{-5}$, i.e., drivers occur 5x more often. In this case, driver mutations arise earlier than in (d) and the tumor becomes more homogeneous.

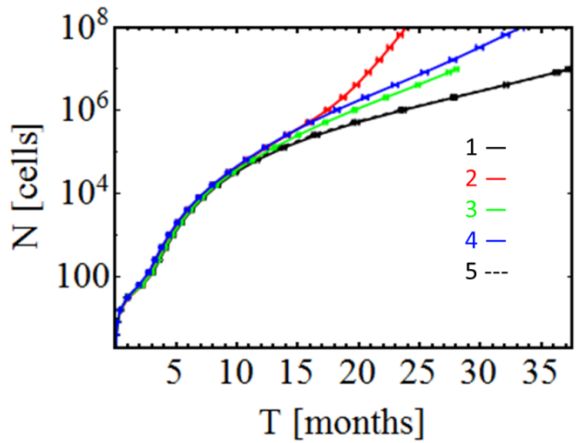

Fig. 13. **Tumor size versus time when drivers affect the dispersal probability.** In all cases, $d=0.9b$, and (1, black) drivers increase the dispersal rate 10-fold ($q=9$) but have no effect on the net growth rate, (2, red) drivers increase both the net growth rate ($s=10\%$) and $M$, (3, green) drivers either (with probability 1/2) increase $M$ 10-fold ($q=9$) or increase the net growth rate by $s=10\%$, (4, blue) drivers increase only the net growth rate by $s=10\%$, (5, black dashed line) neutral case with $M=10^{-7}$, indistinguishable from (1). In all cases (1-3) the initial dispersal probability $M=10^{-7}$.

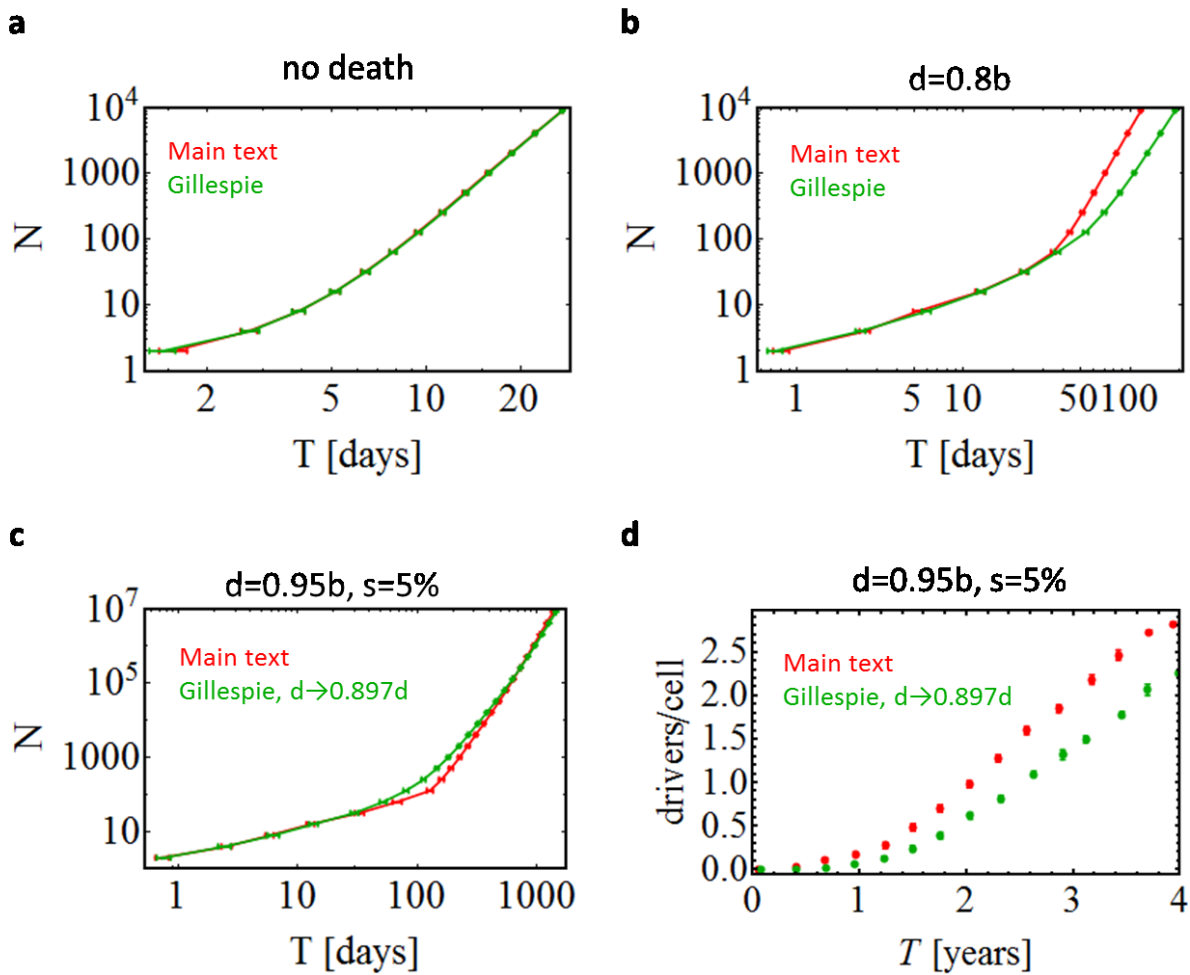

Fig. 14. **Comparison between the algorithm from the main text and the kinetic-Monte Carlo (KMC) algorithm (Supplemental Information).** (**a**) In the absence of death, both models predict identical growth. (**b**) Death makes the model differ for large times. (**c**) The difference almost disappears if the death rate is rescaled as $d \rightarrow 0.897d$ in the KMC algorithm, also in the presence of drivers. (**d**) The number of drivers as a function of time differs slightly between the two models.

| CASE | TUMOR TYPE | EDGE KI % | EDGE Total no. of cells | CENTER KI % | CENTER Total no. of cells | RATIO center:edge |
|---|---|---|---|---|---|---|
| 1 | CHROMOPHOBE RCC | **3.46** | 1013 | **1.11** | 2561 | **0.32** |
| 2 | CHROMOPHOBE RCC | **3.07** | 508 | **1.05** | 938 | **0.34** |
| 3 | CHROMOPHOBE RCC | **2.63** | 524 | **0.44** | 697 | **0.17** |
| 4 | CHROMOPHOBE RCC | **1.58** | 581 | **0.53** | 958 | **0.34** |
| 5 | HCC | **17.14** | 892 | **9.74** | 1637 | **0.57** |
| 6 | HCC | **51.71** | 1079 | **32.84** | 2562 | **0.64** |
| 7 | HCC | **47.37** | 435 | **19.97** | 1397 | **0.42** |
| 8 | HCC | **19.02** | 895 | **13.78** | 1191 | **0.72** |
| 9 | HCC | **15.09** | 1074 | **11.98** | 1094 | **0.79** |
| 10 | HCC | **29.84** | 1305 | **20.87** | 2457 | **0.70** |

Table 1. **Experimental results for the percentage of proliferating cells in the center versus at the edge of solid tumors.** A representative section of each tumor was labeled for the proliferation marker Ki67 (KI), and images of the tumor at the leading edge and the center were acquired as described (Sec. 6.1.2). Proliferation is markedly increased at the leading edge. The average ratio of the number of proliferating cells in the center/at the edge is 0.50 (range 0.17-0.79).

| Parameter | Meaning | Value |
|---|---|---|
| $b$ | Birth rate | $\ln 2 \cong 0.69$ days$^{-1}$ |
| $d$ | Death rate | $0 \dots 0.995 \cdot b$ |
| $s$ | Selective advantage | $0 \dots 0.1$ (=0 … 10%) |
| $\gamma$ | Mutation probability (all GAs) | $0.02$ |
| $\gamma_d$ | Mutation probability (drivers only) | $4 \cdot 10^{-5}$ |
| $\gamma_r$ | Mutation probability (resistance-carrying mutations) | $10^{-7}$ |
| $M$ | Dispersal probability | $0 \dots 10^{-4}$ |

Table 2. **Summary of all parameters used in the model**. If, for a given parameter, many different values have been used in different plots, a range of values used is shown. Birth and death rates can also depend on the number of driver mutations, see Sec. 6.1.



\begin{center}
\large{SUPPLEMENTAL INFORMATION}\\
\vspace{5mm}
\large{Bartlomiej Waclaw, Ivana Bozic, Meredith E. Pittman, Ralph H. Hruban, Bert Vogelstein, and Martin A. Nowak}\\
\end{center}

\section{Computer algorithm}
In each time step, a random cell is chosen, its fate (replication or death) decided, and the time variable updated to account for the elapsed time since the last event. More specifically, for Model A,
\begin{enumerate}
	\item A random tumor cell $i$ is chosen. Let its genotype be $g$.
	\item A nearest neighbor of cell $i$ is picked at random and if it is empty, cell $i$ replicates there and creates a new cell $j$ with probability proportional to the birth rate $b_g$ of cell $i$ divided by the maximal birth rate $b_{\rm max}$ of all cells present in the tumor. This ensures that the replication probability is always less or equal to one, and that it is proportional to the number of empty neighbors of $i$. In the case of no replication (site not empty or rejected), proceed to step 5.
	\item Each of the two cells $i,j$ receive $n_i$ and $n_j$ new mutations, with $n_i,n_j$ drawn independently from the Poisson probability distribution (Eq. (1) in Sec. 6.1). If either $n_i,n_j$ is zero, the genotype of the corresponding cell does not change.
	\item With probability $M$, the new cell $j$ migrates and forms a new ball (microlesion) in the proximity of the surface of the ball from which it originates. The 3d position $\vec{X}_j$ of the new ball is determined as $\vec{X}_j=\vec{X}_i+R_i\vec{x}_i/|\vec{x}_i|$, where $\vec{X}_i$ denotes the position of the ball of cell $i$, $R_i$ is the radius of that ball, and $\vec{x}_i$ is the position of cell $i$ relative to the center of the ball, see Fig. 5a.
	\item Cell $i$ dies with probability equal to its death rate $d_g$ divided by the maximal death rate $d_{\rm max}$ of all existing tumor cells. If all cells within a ball die, the ball is removed. 
	\item Time is updated by a small increment $dt = 1/(b_{\rm max} N)$, where $N$ is the total number of cells in the tumor. This ensures that $b$ has the correct interpretation of probability per unit time, and the number of cells in an exponentially growing population (for example, for large $M$) increases as $N(t)\cong \exp((b-d)t)$.
\end{enumerate}
Therefore, cells replicate only if there is at least one empty site in the neighborhood, but die with rates independent of the number of empty neighbors.
If a cell is completely surrounded by other cells, it cannot replicate. However, in simulations presented in the main text we always assume $d>0$ so that even such ``quiescent'' cells can eventually resume replication if they survive long enough so that one of their neighbors die first and make space for their progeny. We leave the interesting case of $d=0$ (essentially, the Eden model [60] with mutations) to be discussed in future work.

Note that (i) each daughter cell receives different GAs, and that the average number of new GAs per daughter cell per replication is $\gamma/2$ in our notation, (ii) for small $\gamma$ (realistic situation, see Sec. 6.1.1), the probability of a single, new GA in any of the daughter cells is approximately equal to $\gamma$. 

The simulation always begins with a single cell (and thus also a single ball of cells) at position $\vec{x}=(0,0,0)$. We assume that this cell already has a selective advantage over normal cells. We therefore do not model the process of tumor initiation e.g. by inactivation of tumor suppressor genes. The initial cell begins to replicate as described above, and once two or more balls of cells are formed as the result of growth and short-range dispersal, the balls start to repel each other mechanically. This reduces the otherwise imminent and physically impossible overlap between the balls of cells. Because calculating forces and solving equations of motion for hundreds of balls every time a new cell is added to the tumor would be computationally too expensive and required a detailed biophysics model of the tissue, the algorithm we use to reduce the overlap is a less-realistic but more efficient ``shoving'' algorithm similar to the method used in the biofilm simulator iDynomics~[62,63]. The algorithm works by moving apart each pair of overlapping balls along the line joining their centers of mass, see Fig.~5b, and is executed only if the radius $R$ of any ball (measured as the distance from the center of the ball to the most distant cell) has increased by 5\% or more compared to the last time the algorithm was used. A few iterations are usually necessary to relocate all overlapping balls, but because this is done rarely as compared to replication/death of individual cells, it does not have a huge impact on performance for typical scenarios we simulated (a few hundreds of balls). 

For Model B, step (2) of the algorithm is modified as follows: If cell $i$ has no empty neighbors, proceed to step 5. Otherwise, pick up one of them at random and, with probability $b_g/b_{\rm max}$, create a new cell $j$. If no replication, proceed to step 5. Model B differs from A in that the replication probability does not depend on the number of empty (normal) cells in the neighborhood.

For Model C, in step (2) cell $i$ always replicates, regardless of whether there are empty sites nearby or not. The site to which the cell replicates is determined as follows. Ten randomly selected directions out of 26 possible directions are scanned until an empty site is reached, and the direction along which the number of occupied sites is the smallest is chosen. The algorithm proceeds to the first site in this direction, and the scanning is repeated. The whole cycle (scanning/moving by one site) is continued until an empty site is reached. Then each cell encountered during the procedure is shifted to the position of the next site in the chain. This eventually creates an empty site near cell $i$ to which it replicates. This algorithm mimic mechanical displacement of cells towards the surface caused by replication inside the tumor. The path of ``minimal drag'' chosen by the algorithm is self-avoiding, i.e., when choosing the direction for the next step the algorithm checks if a site has been already visited and, if it has, this direction is discarded and another (random) one is chosen. The procedure introduces small artifacts related to lattice symmetries, but deviations from the expected spherical shape of the lesion are negligible, c.f. Fig. 12c.

For Model D, step (5) is modified as follows: cell $i$ dies with probability equal to its death rate $d_g$ multiplied by the fraction of empty nearby sites and divided by $d_{\rm max}$. Thus, both the replication and death rates are proportional to the number of empty neighbors. This causes that cells completely surrounded by other cancer cells not only do not replicate, but also do not die, and form a quiescent core.

\section{\label{sec:param}Model parameters}

{\bf Growth and death rates prior to treatment.} The initial neoplastic/cancer cell has the birth rate $b=\ln 2 \approx 0.69$ days$^{-1}$. This corresponds to a 24h minimum doubling time, though in Model A this is achieved only for cells surrounded by empty space (normal cells), and most cells (surrounded by other tumor cells) will replicate slower on average. 

In the case of metastatic (secondary) lesions, the birth and death rates $b_g=b, d_g=d$ are constant and do not depend on the genotype $g$. The death rate $d$ assumes values between zero and $b$, depending on how aggressive we want the tumor to be. As for primary lesions, in the simulations presented in the main text $b_g=b$ remains the same for all cells prior to treatment and is not affected by mutations. Driver mutations change only the death rate $d_g$ as follows: $d_g=b (1-s)^{k_g}$, where $k_g$ is the number of driver mutations in genotype $g$ including the initial driver mutation (hence $k_g\geq 1$), and $s$ is the selective advantage of each new driver, similarly to Ref.~[8]. In the case of primary lesions, the rate $b$ sets the time scale of the model. There is no independent parameter $d$ for the death rate which is expressed as a fraction of $b$. 

We also consider the case in which $d_g=b(1-s)$ and $b_g = b(1+s)^{k_g-1}$, i.e. the initial cell replicates and dies respectively with rates $b$ and $b(1-s)$, and driver mutations increase the birth rate. As before, the only parameter that sets the time scale here is $b$, the birth rate of the initial cell, and the death rate $d$ is a function of $b$ and $s$. We show below that the results for this case are almost identical to the case from the previous paragraph where drivers reduce the rate of death. Consequently, the model behavior is determined by the difference between the birth and death rates, and it does not matter which of the two rates is affected by driver mutations.

For primary tumors we have explored a range of selective advantages $s=0.5\%\dots 5\%$. Most of the results presented in the main text are for $s=1\%$. For metastatic lesions, where we do not model driver mutations, we take $d/b=0.5\dots 0.9$, with lower death rates corresponding to more agressive and faster growing tumors. This corresponds to the total selective advantage $0.5\dots 0.1=50\%\dots 10\%$ over normal cells.

{\bf Mutation probabilities $\gamma,\gamma_{\rm d},\gamma_{\rm r}$.} These are strictly speaking the average numbers of GAs per parental cell per replication, but since these numbers are small we can think about them as the probabilities of all mutations ($\gamma$), driver mutations ($\gamma_{\rm d}$), and resistant mutations ($\gamma_{\rm r}$). The mutation probabilities are not known precisely and also vary among different cancers. Here we have used numbers that are representative for many cancers. For the total mutation probability we take $\gamma=0.02$. This value corresponds to each of $\sim 6\cdot 10^7$ bp in the diploid exome mutating with probability $\sim 3.3\cdot 10^{-10}$ per bp per generation. For driver mutations, we take $\gamma_{\rm d} = 4\cdot 10^{-5}$ which is about the same as in Ref.~[8]. 
For resistant mutations we assume that $\gamma_{\rm r}=10^{-7}$. This is equivalent, for example, to about 100 potential mutations that confer resistance to any drug targeting a specific protein in cancer cells, each of them occurring with probability $\sim 10^{-9}$, which is an upper estimate on the probability of point mutations in somatic cells [64, 65]. The value of $\gamma_{\rm r}$ is of the same order of magnitude as estimated in Ref.~[66] for targeted EGFR blockade. 

{\bf Dispersal probability $M$.} Since no reliable {\it in vivo} data exists for this parameter, in our model we explore a range of values between $M=10^{-7}$ and $M=10^{-2}$. Our results for the growth time of a metastatic lesion indicate that $M=10^{-7}\dots 10^{-6}$ gives a reasonable agreement with clinically observed tumor growth times, and hence we typically use values from this range. Note that $M$ as defined in our model is the probability that a cell not only detaches from the lesion and moves to a nearby place, but that it also anchors there and starts proliferating. 


\section{Tumor growth rate for the neutral model}
Figures 2, 6 show that in the absence of driver mutations ($s=0$) and for $M=0$ (no dispersal), tumors grow to a roughly spherical shape. The number of cells (and hence also the volume) of the tumor grows approximately as $N\sim T^3$, see Fig. 7. This is easy to understand - the inside of the tumor remains in the state of equilibrium between replication and death, and only the surface is able to expand and invade the surrounding tissue as the result of cancer cells' net growth advantage. Because the number of cells in the surface layer of a spherical tumor is proportional to $N^{2/3}$, the rate of change of the total number of cells ${\rm d}N(t)/{\rm d}t$ is $\sim N(t)^{2/3}$ which gives $N(t)=At^3$. The proportionality coefficient $A$ depends on the thickness of the outermost layer of cells that is not at equilibrium and on the net growth rate of cells in this layer, and in general is a complicated function of $b$ and $d$ which we could not determine analytically.

For $M>0$, however, the model exhibits a transition to exponential growth when the size $N$ becomes much larger than $1/M$, see Fig.~2. This can be explained as follows. Let $N(n,t)$ be the number of balls of cells having $n$ cells at time $t$. Since in our model the balls do not interact with one another (besides mechanical repulsion) and their growth is unaffected by the presence of other balls, we can write the following Master equation for the rate of change of $N(n,t)$ in time:
\bq
	\frac{{\rm d}N(n,t)}{{\rm d}t} = \delta_{n,1} M \sum_{k=1}^\infty B(k) N(k,t) + (1-M)B(n-1)N(n-1,t) - (1-M)B(n)N(n,t), \label{eq:master}
\eq
where $B(n)$ is the rate at which a ball of size $n$ grows to size $n+1$. The first term in Eq.~(\ref{eq:master}) proportional to $M$ accounts for the production of new balls of cells due to dispersal. This happens with the rate equal to the total rate of replication, hence the sum over balls of all sizes $k=1,\dots,\infty$ weighted by $N(k,t)$, the number of balls of size $k$. The second term is the gain term due to replication in a ball of size $n-1$ which creates a ball of size $n$, and the last term is the loss term due to a ball of size $n$ increasing its size by one cell.

If we knew $B(n)$ exactly, we could in principle solve Eq.~(\ref{eq:master}) and find the total size $N(t) \equiv \sum_n n N(n,t)$ as a function of time. Unfortunately, it is not easy to determine $B(n)$ analytically for arbitrary $n$ because of the vast number (and the lack of regularity of) configurations that cells can take in a single ball of size $n\gg 1$. This problem was identified many years ago in the Eden model [60], and it has no known exact solution for large $n$. However, here we are only interested in the asymptotic behavior of $B(n)$ for large sizes $n$ because we anticipate that only this large-$n$ behavior will affect the tumor growth rate at large times. From our earlier results for $M=0$ we conclude that $B(n)\sim n^{2/3}$ for $n\to\infty$ and although finite-size corrections to this formula can be quite strong, we shall neglect them here for the sake of simplicity.

Let us now discuss the large-$t$ solution of Eq.~(\ref{eq:master}). We shall keep a general form of $B(n)$ and replace it by $B(n)\sim n^{2/3}$ in the final step. Equation~(\ref{eq:master}) is a linear differential equation in $N(n,t)$ and therefore its long-time behavior must be dominated by
\bq
	N(n,t) \cong e^{B_{\rm exp} t} p(n), \label{eq:nstat}
\eq
where $B_{\rm exp}$ plays the role of the largest eigenvalue of the linear operator acting on $N(n,t)$ on the r.h.s. of Eq.~(\ref{eq:master}). Its biological interpretation is that of the exponential growth rate of the whole tumor,
\bq
	N(t) = \sum_n n N(n,t) \cong e^{B_{\rm exp} t} \sum_n n p(n),  
\eq
and $p(n)$ is now a time-independent, stationary distribution of ball sizes. One could think that we have just shown that $N(t)$ grows exponentially, but this is not true; it remains to be seen that $B_{\rm exp}={\rm const}>0$ for $M>0$, otherwise sub-leading terms could lead to a sub-exponential (i.e. power-law as we have seen above) growth.

To proceed, we insert Eq.~(\ref{eq:nstat}) into Eq.~(\ref{eq:master}) and obtain the following recursion relation for $p(n)$:
\bq
 B_{\rm exp} p(n) = \delta_{n,1} M \sum_k p(k) B(k) + (1-M)B(n-1)p(n-1) - (1-M)B(n)p(n). \label{eq:pn}
\eq
For $n=1$ this reduces to
\bq
	p(1) = M\frac{\sum_k p(k) B(k)}{B_{\rm exp} + (1-M)B(1)},
\eq
whereas for $n>1$ we can rewrite Eq.~(\ref{eq:pn}) and express $p(n)$ as a function of $p(n-1)$ as follows
\bq
	p(n) = p(n-1) \frac{B(n-1)(1-M)}{B_{\rm exp} + B(n)(1-M)}.
\eq
Iterating the above equation we obtain $p(n)$ as a product of $n$ fractions as in the above equation, times $p(1)$. If we now use this to expand $\sum_k p(k) B(k)$, we have that
\bq
	\sum_k p(k) B(k) = \left[ 1 + \sum_{k=2}^\infty B(k)\left(\prod_{n=2}^k \frac{B(n-1)(1-M)}{B_{\rm exp}+B(n)(1-M)}\right)\right] M\frac{\sum_k p(k) B(k)}{B_{\rm exp} + (1-M)B(1)},
\eq
and after dividing both sides by the common factor $\sum_k p(k) B(k)$ we obtain that $M$ obeys the following equation:
\bq
	\frac{1}{M} = \frac{1}{B_{\rm exp}+B(1)(1-M)} \left[ 1 + \sum_{k=2}^\infty B(k) \left(\prod_{n=2}^k \frac{B(n-1)(1-M)}{B_{\rm exp}+B(n)(1-M)}\right)\right]. \label{eq:bexpm}
\eq
The above equation allows us to calculate $B_{\rm exp}$ as a function of the dispersal probability $M$, albeit in general it must be solved numerically. However, if we now use the asymptotic form $B(n)\sim n^{2/3}$, upon inserting it into Eq.~(\ref{eq:bexpm}) and canceling lower-order terms in $M$ we have
\bq
	B_{\rm exp} \sim M^{1/3} \label{eq:m13}
\eq
for $M\ll 1$. This concludes our proof that the rate of exponential growth is non-zero for $M>0$. Equation (\ref{eq:m13}) also predicts that $B_{\rm exp}$ is proportional to the cubic root of the dispersal probability $M$. Numerical simulations (results not shown) indicate that, for $M>10^{-6}$, the actual $B_{\rm exp}$ increase slower with $M$ than $\sim M^{0.33}$, approximately as $M^{0.25\dots 0.3}$. This is caused by deviations from the law $B(n)\sim n^{2/3}$ for balls smaller than $\sim 10^7$ cells, for which $B(n)$ is better approximated by $\sim n^{0.7\dots 0.75}$.

\section{Reseeding and long-range migration}
In our model each cell that migrates from its original site starts a new ball of cells and there are no interactions (besides mechanical repulsion) between balls of cells. Thus, if we neglect spatial distribution of genetic alterations (GAs) and focus only on the total number of GAs and whole-tumor growth, our algorithm can also model long-range metastasis and reseeding of new lesions.  All our predictions except those related to the spatial distribution of GAs remain valid for the reseeding model. For example, we have shown that, with short-range dispersal, tumor growth becomes exponential with rate $\sim M^{1/3}$, where $M$ is the dispersal probability. If we also include the possibility of reseeding (long-range migration) with probability $R$, the total mass will grow at an increased rate $\sim(M+R)^{1/3}$.

\section{Accumulation of driver mutations}
Figure 11a-c suggests that the number of drivers per cell increases linearly in time. We have made simulations of larger tumors (up to $10^9$ cells), and different $s$, and although the growth rate increases in time, the drivers accumulate approximately at a constant rate (data not shown). This is in contrast to what has been shown in. Ref. [8] for exponentially growing tumors without any spatial structure, where the number of drivers increases exponentially in time. 

This counter-intuitive outcome is the result of the spatial structure of the tumor and all driver mutations having the same selective advantage $s$ in our model. If $s$ is small, a new driver arising in the background of the previous driver mutations spreads within the population with the same speed as the previous drivers, equal to the speed with which the tumor expands into the normal tissue. This expansion, which is similar to travelling fronts in other spatial models, can be demonstrated mathematically. Our statement remains true for sufficiently large tumors in which most cells are inside the ball and not on the surface. Therefore, apart from first few drivers, the $i$-th clonal subpopulation follows the same growth law as the $(i-1)$-th clone but it starts at a later time $t_i$, i.e. its linear extension (``diameter'') $l_i(t)=l(t-t_i)$ where the function $l(t)$ is approximately the same for all clones. Moreover, the times ${t_i}$ at which consecutive driver mutations arise are approximately equally spaced in time. This is because the rate at which new drivers are produced increases very slowly with $i$. This symmetry upon time translation causes linear accumulation of driver mutations over time. We plan to investigate this effect in a subsequent publication.

\section{Alternative simulation method: kinetic Monte Carlo (KMC) algorithm}
Although we talk about rates of different processes, our simulation algorithm is not an exact KMC such as e.g. the Gillespie algorithm. The reason we did not use the standard algorithm was twofold. First, real cells do not reproduce fully stochastically. Regardless of what algorithm is used, a stochastic model will never be fully realistic. 
Second, a fully stochastic simulation would be much slower than our algorithm, even when more advanced methods than the Gillespie algorithm were used.

To check how sensitive are the results of our modelling to the algorithm used, we simulated the model using a KMC, with cells undergoing independent stochastic birth/death events with rates $b$ and $d$. Figure 14 shows that there is only a small difference between the two algorithms. One important drawback (except low speed) of the KMC algorithm is that zero net growth rate does no longer correspond to $d=b$, but to $d\approx 0.897b$ because of a non-trivial dependence of the replication rate on the local density of cells. Consecutively, if $s$ is to be interpreted as the selective advantage, the initial $(b-d)/d$ (the selective advantage of the first driver) used in the KMC algorithm must be multiplied by $0.897$ to make the results comparable to our algorithm. This, and a substantial reduction in the simulation speed, caused that we used the non-KMC algorithm.

\end{document}